\documentstyle[twocolumn,prl,aps]{revtex}
\input epsf

\begin{document}
\draft \twocolumn[\hsize\textwidth\columnwidth\hsize\csname
@twocolumnfalse\endcsname
\title{Effect of Ordering on Spinodal Decomposition of
Liquid-Crystal/Polymer Mixtures}
\author{Amelia M. Lape\~{n}a$^{1}$,  Sharon C. Glotzer$^2$,
Stephen A. Langer$^{3}$, and Andrea J. Liu$^1$}
\medskip
\address{$^1$Department of Chemistry and Biochemistry, UCLA, Los
Angeles, CA 90095-1569}
\address {$^2$Polymers Division and Center for Theoretical and
Computational Materials Science,\\
 National Institute of Standards and Technology, Gaithersburg, MD 20899}
\address {$^3$Information Technology Laboratory, National Institute of
Standards and Technology, Gaithersburg, MD 20899}

\date{\today}
\maketitle

\begin{abstract}
Partially phase-separated liquid-crystal/polymer dispersions display
highly fibrillar domain morphologies that are dramatically different
from the typical structures found in isotropic mixtures.  To explain
this, we numerically explore the coupling between phase ordering and
phase separation kinetics in model two-dimensional fluid mixtures
phase separating into a nematic phase, rich in liquid crystal,
coexisting with an isotropic phase, rich in polymer.  We find that
phase ordering can lead to fibrillar networks of the minority
polymer-rich phase.
\end{abstract}


\vskip2pc]

\narrowtext

Mixtures of liquid crystals with a small amount of polymer
(polymer-stabilized liquid crystals\cite{crawford,mariani,broer}, or
PSLC's) show promise for electro-optic devices such as light shutters
and displays \cite{hikmet1,yang1,yang2,held1}, because the polymer
tends to form a network that aligns the liquid crystal\cite{fung1}.
Since polymers and liquid crystals tend to be immiscible, the
dispersions are prepared by mixing a small amount of miscible monomer
with the liquid crystal and photopolymerizing.  As the polymers grow,
the system phase separates into an ordered phase rich in liquid
crystal and an isotropic phase rich in polymer.  Long before the
system reaches equilibrium, however, the polymerization ``freezes''
the mixture into a crosslinked network of polymer-rich domains.  Thus,
the fabrication of PSLC's involves interplay among three kinetic
processes: polymerization, phase separation, and phase ordering.
Depending on the time scales that control these processes, a rich
variety of morphologies have been
observed\cite{fung2,rajaram,held2,hikmet2}.  Because of the number of
nonequilibrium processes involved, however, there is little
theoretical understanding of the factors that control the domain
morphology. In this Letter, we focus on the interplay between phase
separation (PS) and phase ordering (PO) kinetics in mixtures of short,
rigid polymers ({\em rods}) and long, flexible polymers ({\em coils}),
as a first step towards rational design and control of the network
morphology.

It is well known that thermodynamic factors such as the anisotropy of
the isotropic/nematic interfacial tension can influence domain
morphology, leading to anisotropic domain shapes.  However, there are
also kinetic factors that control domain morphology, such as the
anisotropic diffusion coefficient of a rod.  To capture these
thermodynamic and kinetic effects, we use a Cahn-Hilliard framework
that allows composition and orientational density to evolve in a
coupled fashion as functions of position and time following a
temperature quench\cite{liu96}.  In contrast to earlier studies that
treat orientational density as a scalar order
parameter\cite{dorgan,lansac}, this framework includes the
orientational density's second-order tensorial nature\cite{prost}.
Although it is instructive to study the case of two coupled scalar
order parameters (Model C\cite{hohenberg}), a scalar cannot capture
the direction of nematic order.  Because a vector does not have
head/tail symmetry, it is crucial to retain the tensor order parameter
to obtain domain anisotropy\cite{sagui}.

To assess the effects of phase ordering, we study two systems. The
first (denoted RC) is a mixture of rods and coils. The second (denoted
CC) is a polymer blend, identical to RC except that the rods are
replaced by flexible chains of the same length that do not align.  In
both cases, we solve the linearized coupled partial differential
equations of motion analytically, and the nonlinear equations
numerically into the late-time regime, after quenching the isotropic,
homogeneous mixture into the coexistence region.  We find that phase
ordering dramatically affects morphology, giving rise, for example, to
interconnected networks or elongated domains depending on whether
phase ordering or phase separation is the initially dominant process.

\begin{figure}
\hbox to\hsize{\epsfxsize=1.0\hsize\hfil\epsfbox{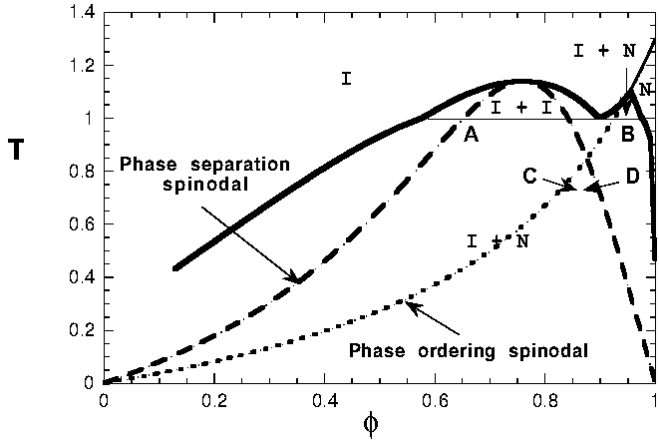}\hfil \medskip}
\caption{Phase diagram of two-dimensional rod/coil blend for the parameters
discussed in the text.  $T$ is the temperature in arbitrary units chosen so
that the IN transition temperature is of order unity, and $\phi$ is the
local area fraction of rods.  The heavy solid line denotes the coexistence
curve.  At low temperatures, an isotropic phase rich in
coils coexists with a nematic phase rich in rods.}
\label{phasediagram}
\end{figure}

We study two-dimensional, incompressible mixtures of chain molecules
made up of monomers of the same size.  In RC, we have short rods that
are $N_{A}=10$ monomers long, and coils that are $N_{B}=100$ monomers
long.  The phase behavior of the system is governed by a bulk free
energy\cite{holyst,liu93} $F_{\rm RC}$ that couples the area fraction
of rods, $\phi$, to the orientational density of the rods, $\tensor
{S}$.  This free energy consists of two parts: the Flory-Huggins free
energy, which governs phase separation, and the Landau-de Gennes free
energy, which governs the isotropic/nematic transition.  The free
energy was calculated within the random phase
approximation\cite{liu93}.  It depends on the rod and coil lengths,
and contains only two free parameters.  These are the two interaction
parameters, namely the Flory parameter, $\chi$, and the Maier-Saupe
parameter, $w$, which control the isotropic and anisotropic
monomer-monomer interactions, respectively.  Both of these parameters
depend on temperature with entropic and enthalpic contributions: $\chi
= \chi_{0} + \chi_{1}/T$ and $w = w_{0} + w_{1}/T$.  We have chosen
the parameters $\chi_{i}$ and $w_{i}$ so that the critical temperature
for phase separation according to the Flory-Huggins free energy,
$T_{c}$, and the isotropic/nematic transition temperature of the pure
rod system, $T_{ni}$, are both of order unity in our (arbitrary)
units, with $T_{ni}$ higher than $T_{c}$.  We use $\chi_{0} = 0.055$,
$\chi_{1} = 0.036$, $w_{0} = 0.55$, and $w_{1} = 0.26$, which yield
$T_{c} = 1.14$ and $T_{ni} = 1.3$.  The resulting RC phase diagram is
shown in Fig.~\ref{phasediagram}.  
Below a triple point, the system
demixes into an isotropic phase rich in coils and a nematic phase rich
in rods.  Above the triple point, there is an isotropic-isotropic
coexistence region and an isotropic-nematic (IN) coexistence
region. The IN coexistence region ends in a critical point that is
connected to the IN transition of a pure rod system by a second-order
transition line.  Note that in three dimensions, the IN transition is
first order, so the phase diagram would be different.  Below the
triple point, however, we find that the IN coexistence curves are
qualitatively identical in two and three dimensions.  The dashed line
represents the spinodal for phase separation (PS), while the dotted
line is the spinodal for phase ordering (PO).

In the CC system, we also use $N_{A}=10$ and $N_{B}=100$ monomers.
Because the ``rods'' are now flexible, the free energy $F_{\rm CC}$
is given by the
Flory-Huggins free energy only, without the Maier-Saupe
contribution \cite{glotzer}.
The parameter $\chi$ is exactly the same as in the RC system.
Thus, CC is identical to RC except that there is no phase ordering.

We study the morphology following four quenches from the isotropic,
homogeneous phase into the IN coexistence region below the triple
point, to the four points marked {\bf A}-{\bf D} in the RC phase diagram.
We use a simplified
version of the equations of motion derived within the dynamical
random phase approximation\cite{liu96}
\begin{eqnarray}
{\partial \phi \over \partial t}& = & \Gamma_{\phi \phi}
\nabla^2{{\delta F(\phi,\tensor {S})} \over \delta \phi} \label{peom}\\
{\partial S^{ij} \over\partial t} & = &
-\Gamma^{ijkl}_{S S}{{\delta F(\phi,\tensor {S})} \over \delta S^{kl}}
\label{Seom}
\end{eqnarray}
where $\Gamma_{\phi \phi}$ and $\Gamma^{ijkl}_{S S}$
are calculated Onsager coefficients that depend on the single chain and
single rod diffusion coefficients; expressions for these coefficients
are provided in Ref.~\cite{liu96}.  The free energy functional
$F(\phi,\tensor {S})$ consists of the bulk free energy and a nonlocal
free energy that controls the cost of gradients in composition and
orientational density\cite{liu93}.  Note that the cost of gradients in
orientational density are determined by the Frank elastic constants
that characterize nematic elasticity.
For CC, the time evolution
is given by Eq.~\ref{peom} alone, since $\tensor S = 0$ and $\delta
F_{\rm CC}/\delta S^{ij} = 0$.
We emphasize
that for both systems, the equations of motion are not
phenomenological; they are derived from microscopic
models. We solve the discretized equations on
a square lattice of size $150^{2}$ using a variable timestep
Runge-Kutta numerical integration scheme.  We have also varied the
number of lattice points and the size of the mesh to verify that our
observations do not depend on system size or discretization.

\begin{figure}
\hbox to\hsize{\epsfxsize=0.9\hsize\hfil\epsfbox{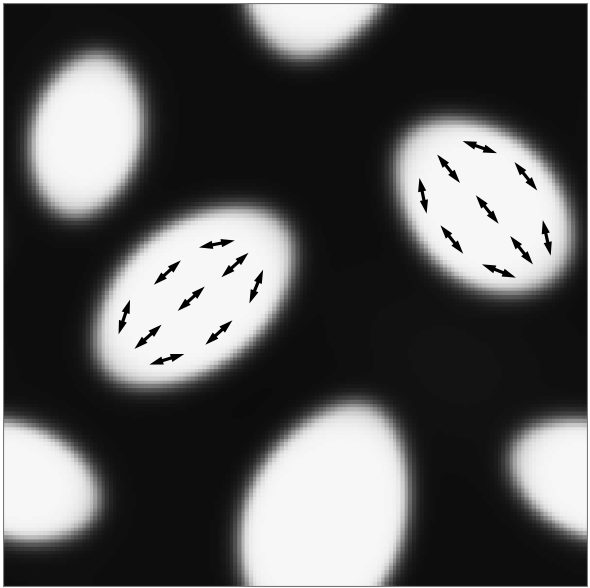}\hfil \medskip}
\caption{Late-time snapshot of the local composition for the RC system
when quenched to {\bf A}, as indicated in Fig.~\ref{phasediagram}.
We use a gray scale where black
corresponds to pure coils and white corresponds to pure rods.
Arrows represent the local nematic director (shown for two droplets only).}
\label{quenchA}
\end{figure}

{\it Quench {\bf A}}: Here we quench to a point where RC is initially
unstable with respect to PS, but metastable with respect to PO.  Thus,
the point marked {\bf A} in Fig.~\ref{phasediagram} lies below the
spinodal for PS and above the one for PO.  For CC, where phase
ordering is absent, this quench leads to circular droplets rich in the
shorter coils, in a matrix rich in the longer coils.  RC initially
displays almost identical behavior with circular droplets rich in
rods.  Once the concentration of rods in the droplets is comparable to
the concentration at the IN spinodal, however, the rods in the
droplets begin to order.  At this time, the droplets abruptly expel
more coils because the coils are less soluble when the rods are
ordered.  In addition, the droplets must develop defects because the
rods want to be parallel to each other and parallel to the droplet
interface.  As the rods order, a pair of defects forms inside each
droplet and separates, with the two defects moving along the director
in opposite directions towards the edge of the droplet.  The magnitude
of orientational order is lower at the edges than in the center; thus,
the defects migrate towards the edges to lower their energy.  Since
the overall system is isotropic, the long axis of each droplet is
randomly oriented.  Note that it is essential to retain the tensorial
nature of the orientational order parameter in order to obtain
randomly-oriented elongated domains.  Fig.~\ref{quenchA} shows a
snapshot of the RC system at a time at which there is significant
phase separation and ordering.

{\it Quench {\bf B}}:
For this quench, CC
is metastable with respect to PS, so isolated droplets rich in the
long coils would form
by nucleation if we had included thermal noise.
Since we have not included thermal noise, the CC system does not evolve.
In this region RC is also initially metastable with respect
to PS, but it is unstable with respect to PO. The
instability towards orientational ordering eventually drives the system to
phase separate because the two order parameters are coupled.  This
makes physical sense:  once the rods are strongly aligned, they expel the
coils into isolated droplets.  Although there is no
orientational order within the coil-rich droplets, the
droplets are anisotropic because of the nematic elasticity
in the surrounding rod-rich matrix.

The appearance of droplets even when the system is only metastable to
phase separation is reminiscent of recent theoretical work on protein
and polymer crystallization, which shows that a quench below the
spinodal for one order parameter (density) can promote domain growth in another
order parameter (crystallization)\cite{frenkel,olmsted}.  The advantage
of our system is that we can
calculate the coupled equations of motion for our two order parameters
>from a microscopic model of rods and coils.

\begin{figure}
\hbox to\hsize{\epsfxsize=0.90\hsize\hfil\epsfbox{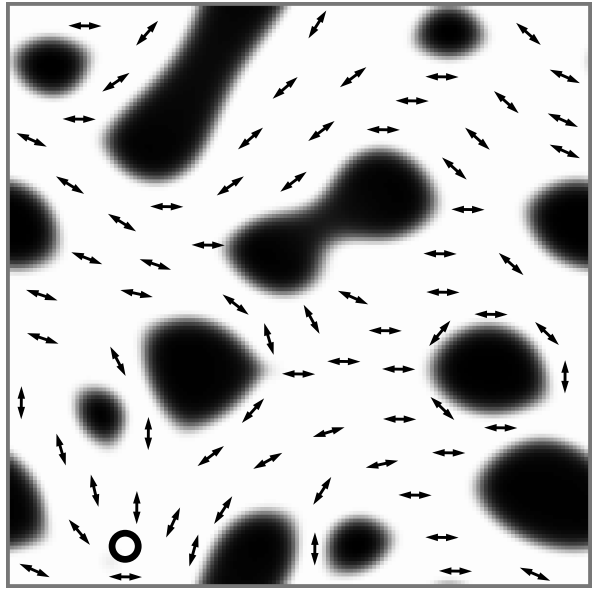}\hfil \medskip}
\caption{Snapshot of the local composition for the RC system
at a late time following a quench to {\bf C}.  Black
corresponds to pure coils and white corresponds to pure rods.  Arrows
illustrate the local nematic director, and a defect is marked with a
circle.}
\label{quenchC}
\end{figure}

{\it Quench {\bf C}}: In this quench, RC is initially unstable both to
PS and to PO, but is more unstable with respect to PS.  In both RC and
CC, the system forms a bicontinuous network that rapidly breaks up
into droplets rich in long coils.  In RC, however, defects in the
surrounding rod-rich domains give rise to droplets that are
noncircular, as shown in Fig.~\ref{quenchC}.  Furthermore, the
coupling of PS and PO leads to a faster onset of phase separation in
RC than in CC, because the rod-rich regions tend to expel coils as the
rods order.

\begin{figure}
\hbox to\hsize{\epsfxsize=0.9\hsize\hfil\epsfbox{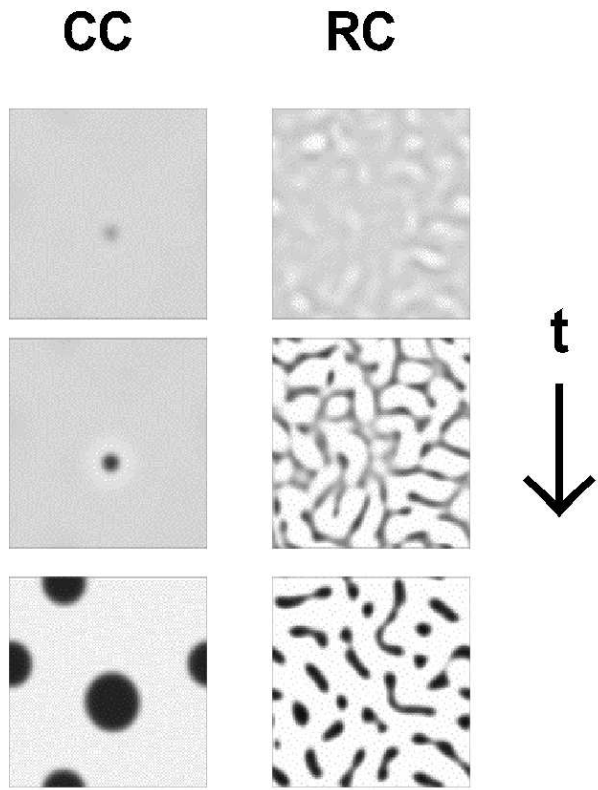}\hfil \medskip}
\caption{Snapshots of local composition as a function of time $t$ in
CC (left) and RC (right) following a quench to point {\bf D} on
the phase diagram in Fig.~\ref{phasediagram}.  We use a gray scale where
black corresponds to pure long coils in CC and to pure coils in RC,
while white corresponds to pure short coils in CC and pure rods in
RC.  Adjacent snapshots for the two systems have comparable degrees of
phase separation.  Note that phase ordering in RC drives the system
to a fibrillar network morphology of coil-rich domains at intermediate
times.}
\label{quenchD}
\end{figure}

{\it Quench {\bf D}}: As in Quench {\bf C}, RC is initially unstable
both to PS and to PO.  However, in this quench it is more unstable
with respect to PO.  Once the degree of order is significant, then
phase separation begins.  Because of PO, PS begins much earlier in RC
than in CC; by the time CC displays any noticeable phase separation,
the concentration difference between the two phases in RC has already
reached its equilibrium value.  Fig.~\ref{quenchD} presents snapshots
of the systems as they evolve in time.  In CC, an interconnected
network rich in the longer coils initially appears.  However, it
quickly breaks up into droplets that become increasingly circular over
time.  In RC, on the other hand, small highly-ordered rod-rich
droplets initially form in a coil-rich matrix.  As these rod-rich
droplets grow, the surrounding coil-rich region shrinks into a network
of interconnected domains that are strikingly fibrillar, despite the
fact that the coil-rich phase is the minority phase.  This is in
contrast to systems with only a compositional order parameter (Model B
\cite{hohenberg}), where only the majority phase can form networks.
Eventually, this network breaks up to form coil-rich droplets; again,
these are noncircular because of defects in the surrounding rod-rich
regions. However, in experiments the crosslinking of the polymer
network may arrest the phase separation at the fibrillar network
stage.

In summary, we find that phase ordering can significantly influence
domain morphology, especially when (1)  the system is unstable to both phase
separation and phase ordering and (2)  orientational order is significant
at the onset of phase separation.  This is consistent with morphologies
observed in PSLC's, where fibrillar networks of the minority
polymer-rich domains
form when the system is initially ordered\cite{fung2,rajaram,held2}.
Our results show
that we can learn much about domain morphology from the phase
diagram and spinodal lines, and that the coupling of phase separation
and ordering kinetics leads to rich phenomena that warrant further
study.

We thank Abdullah Al Sunaidi, Rashmi Desai, Weinan E, Peter Palffy-Muhoray,
and Rebecca M. Nyquist for
stimulating and instructive discussions that led to this research.
This work was supported by NIST through the Center for Theoretical and
Computational Materials Science and by the NSF through Grant No.
CHE-9624090 (AJL).

\end{document}